\documentclass[aps,prd,notitlepage,nofootinbib,english,showpacs,10pt]{revtex4-1}

\usepackage[latin1]{inputenc}
\usepackage{float}
\usepackage{latexsym}
\usepackage{graphicx}
\usepackage{amssymb}
\usepackage{amsmath}
\usepackage{amsfonts}
\usepackage{hyperref}
\usepackage{color}
\usepackage{cool}
\usepackage{subfig}
\usepackage{dcolumn}

\begin{document}

\title{Testing the interaction between dark energy and dark matter with Planck data}

\author{Andr\'e A. Costa$^a$}\email{alencar@if.usp.br}
\author{Xiao-Dong Xu$^b$}\email{ammonitex@163.com}
\author{Bin Wang$^b$}\email{wang\_b@sjtu.edu.cn}
\author{Elisa G. M. Ferreira$^{a,c}$}\email{elisa@if.usp.br}
\author{E. Abdalla$^a$}\email{eabdalla@usp.br}
\affiliation{$^a$Instituto de F\'isica, Universidade de S\~ao Paulo, C.P.
  66318, 05315-970, S\~ao Paulo, SP, Brazil}
\affiliation{$^b$Department of Physics and Astronomy, Shanghai Jiao Tong University, 200240 Shanghai, China}
\affiliation{$^c$Department of Physics, McGill University, Montr\'eal, QC, H3A 2T8, Canada}
\date{\today}

\begin{abstract}

Interacting Dark Energy and Dark Matter is used to go beyond the standard cosmology. We base our arguments on
Planck data and conclude that an interaction is compatible with the observations and can provide a
strong argument towards consistency of different values of cosmological parameters.

\end{abstract}

\maketitle

\section{Introduction}

The incredible amount of precise astronomical
data released in the past few years provided
great opportunities to answer problems in
cosmology and astrophysics. Recently, the Planck
team released their first data with higher
precision and new full sky measurements of the
Cosmic Microwave Background (CMB) temperature
anisotropies in a wide range of multipoles
($l<2500$) \cite{Ade:2013ktc,
Ade:2013zuv,Planck:2013kta}. Such a precision
allows us to test cosmological models  and
determine cosmological parameters with a high
accuracy.

The Planck team analysis showed that the universe
is flat and in full agreement with the
$\Lambda$CDM cosmological model, especially for
the high multipoles ($l>40$). However, the value
of the Hubble parameter today presents about
$2.5\sigma$ tension in comparison with other low
redshift probes, for example the direct
measurement done by Hubble Space Telescope (HST)
\cite{Riess:2011yx}. If this difference is not
introduced by systematics, this can point out to
an observational challenge for the standard
$\Lambda$CDM model. The Planck determination of
$H_{0}$ assumed a theoretical $\Lambda$CDM model,
which can influence its value on $H_{0}$.

Theoretically the $\Lambda$CDM model itself is
facing challenges, such as the cosmological
constant problem\cite{Weinberg:1988cp} and the
coincidence problem\cite{Chimento:2003iea}. The first
problem refers to the small observed value of the
cosmological constant incompatible with the
vacuum energy description in field theory.
The second problem refers to the fact that we
have no natural explanation why the energy
densities of dark matter and vacuum energy are of
the same order today. These problems open the
avenue for alternative models of dark energy to
substitute the cosmological constant description. For
example, the use of a component with dynamically
varying equation of state parameter to describe the dark
energy. However, although it can alleviate the coincidence
problem, it suffers the fine tuning problem. Thus
these models are not prevailing.

Another way to alleviate the coincidence problem,
which embarrasses the standard $\Lambda$CDM
cosmology is to consider an interaction between
dark energy and dark matter. Considering that
dark energy and dark matter contribute
significant fractions of the contents of the
universe, it is natural, in the framework of
field theory, to consider an interaction between
them. The appropriate interaction can accommodate
an effective dark energy equation of state in the
phantom region at the present time. The
interaction between dark energy and dark matter
will affect significantly the expansion history
of the Universe and the evolution of density
perturbations, changing their growth. The
possibility of the interaction between dark
sectors has been widely discussed in the
literature \cite{Amendola:1999er,Amendola:2003eq,Amendola:2005ps,
Pavon:2005yx,delCampo:2008sr,Boehmer:2008av,Chen:2008ca,
Olivares:2005tb,Olivares:2007rt,Valiviita:2008iv,He:2008si,
Corasaniti:2008kx,Jackson:2009mz,Pavon:2007gt,Wang:2007ak,
Wang:2006qw,Simpson:2010yt,Zimdahl:2005bk,Guo:2007zk,
Feng:2008fx,Valiviita:2009nu,Xia:2009zzb,He:2009pd,Martinelli:2010rt,
Honorez:2010rr,He:2008tn,He:2009mz,CalderaCabral:2009ja,
He:2010ta,Bertolami:2007zm,Bertolami:2007tq,Abdalla:2007rd,
Abdalla:2009mt,Pellicer:2011mw}.
Determining the existence of dark matter and dark
energy interactions is an observational endeavor
that could provide an interesting insight into
the nature of the dark sectors.

Since the physical properties of dark matter and
dark energy at the present moment are unknown, we
cannot derive the precise form of the interaction
from first principles. For simplicity, most
considerations of the interaction in the
literature are from phenomenology. Attempts to
describe the interaction  from  field theory have
been proposed in
\cite{Pavan:2011xn,Micheletti:2009pk,Micheletti:2009jy}.
In this paper we will concentrate on a
phenomenological model of the interaction between
dark matter and dark energy, which is in a linear
combination of energy densities of the dark
sectors $Q_c = 3H(\xi_1\rho_c + \xi_2\rho_d)$
\cite{He:2008si,He:2009mz,Koyama:2009gd},
where $\xi_1$ and $\xi_2$ are dimensionless
parameters and assumed to be time independent for
simplicity. This model was widely studied in
\cite{He:2008si,Wang:2006qw,Zhou:2008hf,Abdalla:2007rd,
Wang:2005ph,Wang:2005jx,He:2010im}.
It was disclosed that the interaction between
dark matter and dark energy influences the CMB at
low multipoles by the late integrated Sachs-Wolf
(ISW) effect \cite{He:2009pd,Honorez:2010rr} and at
high multipoles through gravitational lensing
\cite{He:2010im,Salvatelli:2013wra}. With the WMAP
data \cite{He:2009pd,Honorez:2010rr} together with
galaxy clusters observations \cite{Abdalla:2007rd,Abdalla:2009mt} and also recent kinetic
Sunyaev-Zel'dovich effect observations
\cite{Xu:2013jma}, it was found that this
phenomenological interaction between dark energy
and dark matter is viable and the coupling
constant is positive indicating that there is
energy flow from dark energy to dark matter,
which is required to alleviate the coincidence
problem and to satisfy the second law of
thermodynamics \cite{Pavon:2007gt}.

It is of great interest to employ the latest high
precision Planck data to further constrain the
phenomenological interaction model. This is the
main motivation of the present work. We will
compare the constraint from the Planck data with
previous constraints from WMAP data \cite{He:2009pd,Honorez:2010rr}. 
Especially, we want to examine
whether, with the interaction between dark matter
and dark energy, we can reduce the tension on the
value of $H_0$ at present. We will combine the
CMB data from Planck with other cosmological
probes such as the Baryonic Acoustic Oscillations
(BAO), Supernovas and the latest constraint on
the Hubble constant \cite{Riess:2011yx}. We want
to see how these different probes will influence
the cosmological parameters and put tight
constraints on the interaction between dark
sectors.

This paper is organized as follows: in Section
\ref{sec:coupled_models} we will describe the
phenomenological interaction model between dark
sectors and present the linear perturbation
equations. In Section \ref{sec:analysis} we will
explain the methods used in the analysis. Section
\ref{sec:results} will present the results of the
analysis and discussions. In the last section we
will summarize our results.

\section{The phenomenological model on the interaction between dark sectors}

\label{sec:coupled_models}

The formalism describing the evolution of matter
and dark energy density perturbations without
\cite{Kodama:1985bj,Ma:1995ey}  and with dark matter and
dark energy interaction \cite{He:2009mz} is
well established. If dark matter and dark energy
are coupled with each other, the energy-momentum
tensor $T^{\mu\nu}_{(\lambda)}$ of each individual
component $\lambda=c,d$ is no longer conserved.
Instead,
\begin{equation}
\nabla _{\mu} T^{\mu\nu}_{(\lambda)} = Q^{\nu}_{(\lambda)}\,,
\label{EE_int}
\end{equation}
where $Q^{\nu}_{(\lambda)}$ is the four vector
governing the energy-momentum transfer between
dark components and the subscript $(\lambda)$ can
refer to dark matter $(c)$  and dark energy
$(d)$, respectively. With interaction between
dark sectors, dark matter and dark energy
components are not conserved separately, but the
energy-momentum tensor of the whole dark sector
is still conserved, thus, $Q^{\nu}_{(c)}=
-Q^{\nu}_{(d)}$.

Assuming spatially flat
Friedmann-Robertson-Walker background, from the
energy conservation of the full energy-momentum
tensor, we can derive the equations of evolution
of the mean dark matter and dark energy densities
\begin{alignat}{2}
\dot{\rho}_{c}+3\mathcal{H}\rho_{c}= & a^2Q^{0}_{c}= & +aQ\,,\nonumber \\
\dot{\rho}_{d}+3\mathcal{H}\left(1+\omega\right)\rho_{d}= & a^2Q^{0}_{d}= & -aQ\,,\label{eq:intera_fen}
\end{alignat}
where the derivatives and the Hubble parameter
$\mathcal{H}$ are in conformal time, $\rho_{c}$
is the energy density for dark matter,
$\omega=p_{d}/\rho_{d}$ is the equation of state
of dark energy, $a$ is the scale factor and $Q$
was chosen to be the energy transfer in
cosmic time coordinates.  We emphasize that
the homogeneity and isotropy of the background
require the spatial components of
$Q^{\nu}_{(\lambda)}$ to be zero.

We concentrate on the phenomenological
interaction as a linear combination of energy
densities of dark sectors with the form of
$Q=3H(\xi_1\rho_c + \xi_2\rho_d)$, which
describes the energy transfer. In the above
expression of the continuity equations, if $Q>0$,
we have the dark energy transfers energy to the
dark matter while if it is negative, the transfer
is in the opposite direction. In studying the
curvature perturbation it has been made clear
that when the interaction is proportional to the
energy density of dark energy ($Q =
3H\xi_2\rho_d$), we get a stable curvature
perturbation except for $\omega=-1$;
however, when the interaction is proportional to
the dark matter density ($Q = 3H\xi_1\rho_c$) or
total dark sectors ($Q = 3H\xi(\rho_c+\rho_d)$),
the curvature perturbation can only be stable
when the constant dark energy equation of state
satisfies $\omega < -1$ \cite{He:2008si}. For the
case of a time-dependent dark energy equation of
state, the stability of curvature perturbations
was discussed in
\cite{Corasaniti:2008kx,Jackson:2009mz}. With the
interaction, the effective background equations
of state for the dark matter and dark energy
change to
\begin{align}
\omega_{c, eff}=-\frac{a^2Q^{0}_{c}}{3\mathcal{H}\rho_{c}}\,, \qquad \omega_{d,eff}=\omega - \frac{a^2Q^{0}_{d}}{3\mathcal{H}\rho_{d}}\,,
\end{align}
where $\omega$ is the equation of state of dark
energy. We summarize different forms of the
interaction with the effective background
equation of state in Table \ref{models} as done in
\cite{He:2010im}, we label our models with Roman
numbers.

\begin{figure}[H]
\centering \subfloat[
]{\includegraphics[scale=0.45]{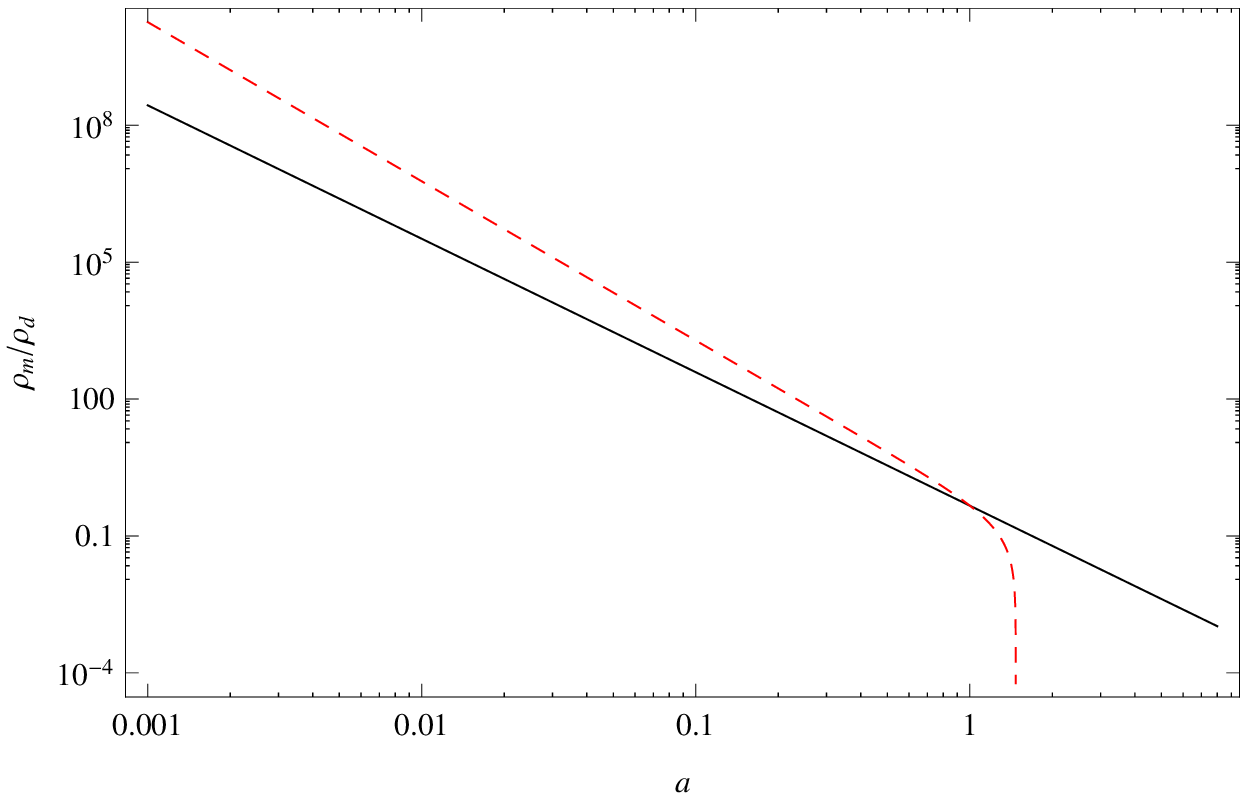}}
\subfloat[
]{\includegraphics[scale=0.45]{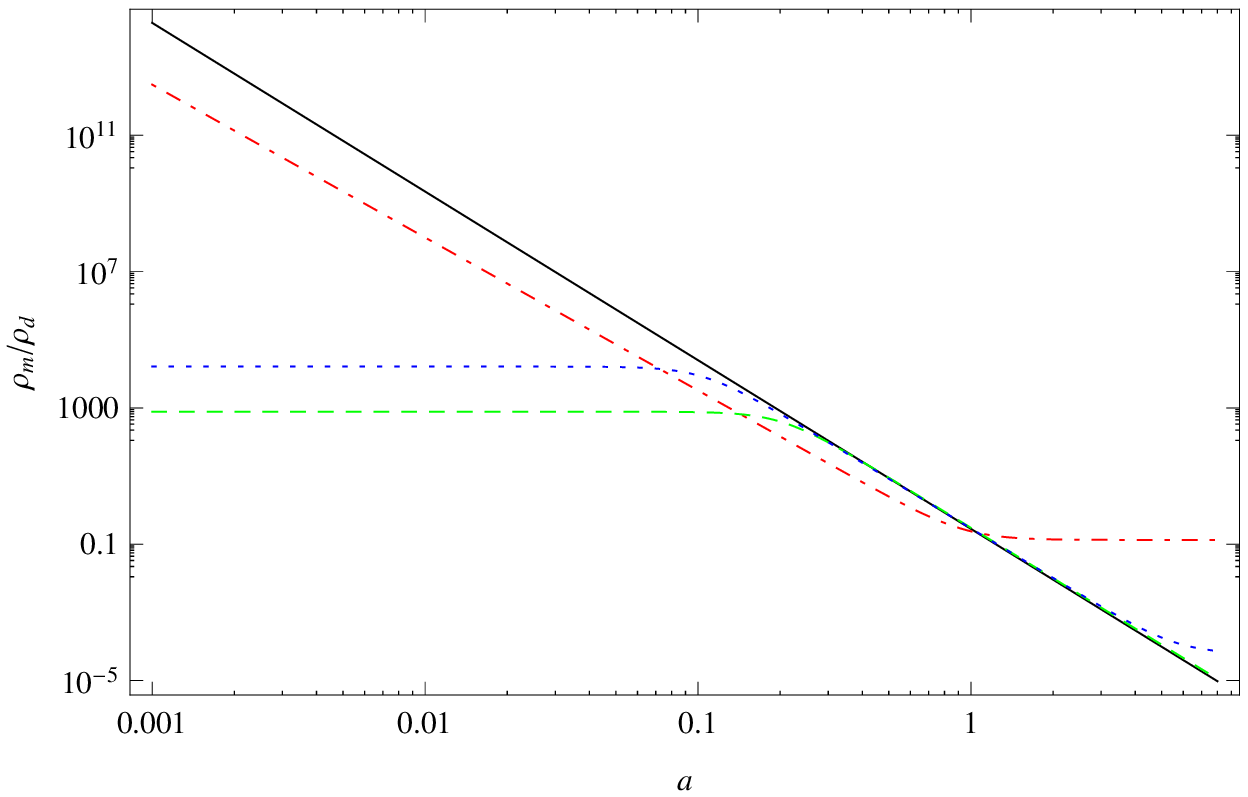}}
\caption{(Color online). Evolution of the dark
energy/dark matter energy density ratio $r\equiv
\rho_c/\rho_d$ in a model with $Q =
3H(\xi_1\rho_c + \xi_2\rho_d)$ for different
coupling constants. (a) The red dashed line corresponds to Planck bestfit Model I, with $\xi_2 = -0.1881$ corresponding to the lowest value in the 68\% C.L. as in Table \ref{bestfit1}. The black solid line has the same parameters but no interaction. (b) The black solid line corresponds to a non-interacting model with $w=-1.65$ and $\Omega_d=0.78$. The red dot-dashed line describes Model II listed in the first column of Table \ref{bestfit2} with $\xi_2=0.2$. The green dashed line corresponds to Planck bestfit Model III (see Table \ref{bestfit3}); and blue dotted line to Planck bestfit Model IV (see Table \ref{bestfit4}).} \label{ratio_rho}
\end{figure}

In order to solve the coincidence problem, we
require the ratio of the energy densities of dark
matter and dark energy, $r=\rho_{c}/\rho_{d}$, to
be a constant in the expansion history of our
universe. This leads to a quadratic equation,
\begin{equation}
    \xi_{1} r^{2}+\left( \xi_{1}+\xi_{2}+\omega \right) r+\xi_{2}=0.
    \label{r_eq}
\end{equation}
The solutions of this equation can lead to unphysical results, as negative energy density of cold DM in the past or complex roots. For different phenomenological models of the
interaction between dark sectors, the conditions
to obtain physical results, positive energy densities and real roots, were summarized in \cite{He:2010im} as shown in Table \ref{models}. Fig.\ref{ratio_rho} illustrates the behavior of $r$ for the four interacting models. We observe that, for the interaction proportional to the energy density of dark energy, a positive interaction can help to alleviate the coincidence problem as there is a longer period for the energy densities of dark matter and dark energy to be comparable. In contrast, a negative interaction can not alleviate the coincidence problem. For the interaction proportional to the energy density of dark matter or to the sum of both energies, the ratio $r$ presents a scaling behavior.

\begin{table}[htb]
\centering
\caption{In this table we present the different coupling models considered with its constraints, dark energy equation of state and the effective equation of state for both fluids.}
\begin{tabular}{|c|c|c|c|c|c|}
 \toprule
    Model & Q & DE EoS & $\omega_{c,eff}$ & $\omega _{d,eff}$ & Constraints \\
    \hline
    I & $3\xi _{2} H\rho _{d}$ & $-1 <  \omega < 0$ & $-\xi _{2} /r$ & $\omega + \xi _{2}$ & $\xi _{2} < -2 \omega \Omega _{c}$ \\
    \hline
    II & $3\xi_{2} H\rho_{d}$  & $\omega < -1$ & $-\xi_{2}/r$ & $\omega + \xi_{2}$ & $\xi_{2} < -2\omega \Omega_{c}$ \\
\hline
    III & $3\xi_{1} H\rho_{c}$  & $\omega < -1$ & $-\xi_{1}$ & $\omega+\xi_{1}r$ & $0 < \xi_{1} < -\omega /4$ \\
    \hline
    IV & $3\xi H \left(\rho_{d} + \rho_{c} \right) $ & $\omega < -1$ & $-\xi \left( 1+1/r \right)$ & $\omega+\xi \left(r+1\right)$ & $0 < \xi < -\omega /4$ \\
    \lasthline
\end{tabular}
\label{models}
\end{table}

From the background dynamics we see that when we introduce
the phenomenological interaction between dark
sectors, it is possible to have the scaling
solution of the ratio between dark matter and
dark energy, which can help to alleviate the
coincidence problem. However, in the background
dynamics there appears an inevitable degeneracy
between the coupling in dark sectors and the
dark energy equation of state. In general this
degeneracy cannot be broken by just investigating
the dynamics of the background spacetime, except
in the case when the coupling is proportional to
the dark matter density (Model III) as was
discussed in \cite{He:2010im}. It is expected that the degeneracy between the
coupling and other cosmological parameters can be
solved in the perturbed spacetime by considering
the evolution of the perturbations of dark energy
and dark matter.  The perturbed FRW space-time
has a metric given by
\begin{equation}\label{dse}
ds^2 = a^2\left[-(1 + 2\psi)d\tau^2 + 2\partial_i B d\tau dx^i + (1 + 2\phi)\delta_{ij}dx^idx^j + D_{ij}Edx^idx^j\right],
\end{equation}
where
\begin{equation}
D_{ij} = \left(\partial_i\partial_j - \frac{1}{3}\delta_{ij}\nabla^2\right).
\end{equation}
The functions $\psi$, $B$, $\phi$ and $E$
represent the scalar metric perturbations. In the
synchronous gauge $\psi = B = 0$.

We will use an energy-momentum tensor of the form
\begin{equation}
T^{\mu\nu}(\tau,x,y,z) = (\rho + P)U^\mu U^\nu + Pg^{\mu\nu},
\end{equation}
where $\rho$, $P$ are composed by a term
depending only on time plus a small perturbation
that depends on all coordinates. The
four-velocity reads
\begin{equation}
U^\mu = a^{-1}(1 - \psi, \vec{v}_{(\lambda)}),
\end{equation}
where $\vec{v}_{(\lambda)}$ can be written as
minus the gradient of a peculiar velocity
potential $v_{(\lambda)}$ plus a zero divergence
vector. Only the first one contributes to scalar
perturbations. In the Fourier space, we use the
convention to divide the velocity potential by an
additional factor of $k \equiv |\vec{k}|$ so that
it has the same dimension as the vector part.
Thus,
\begin{equation}
\theta \equiv \nabla \cdot \vec{v} = - \nabla^2 v = kv.
\end{equation}

Following \cite{Valiviita:2008iv} we write the
perturbed pressure of dark energy as
\begin{equation}
\delta P_{d} = c_e^2\delta_{d}\rho_{d} + (c_e^2 - c_a^2)\left[\frac{3{\cal H}(1 + \omega)v_{d}\rho_{d}}{k} - a^2Q_{d}^0\frac{v_{d}}{k}\right],
\end{equation}
where $\delta=\delta \rho / \rho$ is the density
contrast, $c_e^2$ is the effective sound speed of
dark energy at its rest frame, which we set to
one, and $c_a^2$ is the adiabatic sound speed. As
discussed in \cite{He:2010im}, the perturbed four
vector $\delta Q^{\nu}_{(\lambda)}$ can be
decomposed into
\begin{equation}
\delta Q^{0}_{(\lambda)} = \pm \left(-\frac{\psi}{a}Q + \frac{1}{a}\delta Q\right)\,, \qquad \delta Q_{p (\lambda)} = \left. Q_{p (\lambda)}^I \right|_t + Q^{0}_{(\lambda)}v_t.
\end{equation}
Here the $\pm$ sign refers to dark matter or dark
energy respectively, and $\delta Q_{p (\lambda)}$
is the potential of the perturbed energy-momentum
transfer $\delta Q^{i}_{(\lambda)}$. $\left. Q_{p
(\lambda)}^I \right|_t$  is the external
non-gravitational force density and $v_t$ is the
average velocity of the energy transfer. In this
paper we consider that there is no
non-gravitational interaction between dark energy
and dark matter, only inertial drag effect
appears due to stationary energy transfer. Thus
$\left. Q_{p (\lambda)}^I \right|_t$ and $v_t$
vanish which implies that $\delta
Q^{i}_{(\lambda)}=0$.

In the synchronous gauge, the
linear order perturbation equations for dark
matter and dark energy read \cite{He:2010im}
\begin{align}
     \dot{\delta}_{c} & =  -(kv_{c} + \frac{\dot{h}}{2}) + 3\mathcal{H}\xi_{2} \frac{1}{r} \left( \delta_{d}-\delta_{c} \right)\,, \label{linear_pert_1} \\
     \dot{\delta}_{d} & = -\left(1+\omega \right) (k v_{d} + \frac{\dot{h}}{2})+3\mathcal{H} (\omega - c_{e}^{2}) \delta_{d}+3\mathcal{H} \xi_{1} r \left( \delta _{d}-\delta _{c} \right)  \nonumber \\
                            & -3\mathcal{H} \left( c_{e}^{2}-c_{a}^{2} \right) \left[ 3 \mathcal{H} \left( 1+\omega \right) + 3\mathcal{H} \left( \xi_{1} r+\xi_{2} \right) \right]\frac{v_{d}}{k}      \,, \label{linear_pert_2} \\
     \dot{v}_{c} & = -\mathcal{H}v_{c} -3\mathcal{H}(\xi_{1} + \frac{1}{r}\xi_{2})v_{c} \,, \label{linear_pert_3} \\
     \dot{v}_{d} & = -\mathcal{H}\left(1-3c_{e}^{2} \right) v_{d}+\frac{3\mathcal{H}}{1+\omega} \left( 1+c_{e}^{2} \right) \left(\xi_{1} r+\xi_{2} \right) v_{d}+ \frac{kc_{e}^{2}\delta _{d}}{1+\omega}\,,
     \label{linear_pert_4}
\end{align}
where $h = 6\phi$ is the synchronous gauge metric
perturbation and $v_{d}$ is the peculiar velocity
of the dark energy. The peculiar velocity of the
dark matter $v_{c}$ is considered to be null
because we are working in a frame comoving with
the matter fluid. To solve equations
(\ref{linear_pert_1}, \ref{linear_pert_2}, \ref{linear_pert_3}, \ref{linear_pert_4})
we set initial conditions according to
\cite{He:2008si}. In the linear perturbation
formalism, the influence of the interaction
between dark energy and dark matter on the CMB
can be calculated by modifying the CAMB code
\cite{Lewis:1999bs}. This can be done by directly
including equations (\ref{eq:intera_fen},
\ref{linear_pert_1}, \ref{linear_pert_2},
\ref{linear_pert_3} and \ref{linear_pert_4}) in
the code.

In \cite{He:2010im}, it was uncovered that in
addition to modifying the CMB spectrum at small
$l$, the coupling between dark sectors can shift
the acoustic peaks at large multipoles. While the
change of equation of state of dark energy can
only modify the low $l$ CMB power spectrum, it
leaves the acoustic peaks basically unchanged.
This provides the possibility to break the
degeneracy between the coupling and the equation
of state of dark energy in the linear
perturbation theory. Furthermore, it was observed
that the abundance of dark matter can influence
the acoustic peaks in CMB, especially the first
and the second ones. The degeneracy between the
abundance of the dark matter and the coupling
between dark sectors can be broken by examining
the CMB spectrum at large scale, since only the
coupling between dark sectors influences the
large scale CMB spectrum. Theoretically it was
observed that there are possible ways to break
the degeneracy between the interaction, dark
energy equation of state and the dark matter
abundance in the perturbation theory
\cite{He:2010im}. This can help to get tight
constraint on the interaction between dark energy
and dark matter.

In the following we are going to extract the
signature of the interaction and constraints on
other cosmological parameters by using the Planck
CMB data together with other observational data
and compare with previous results obtained in
\cite{He:2010im} by employing WMAP data.

\section{Method on data analysis}
\label{sec:analysis}

We compute the CMB power spectrum with the
modified version of CAMB code
\cite{Lewis:1999bs}, in which we have included
both background and linear perturbation equations
in the presence of a coupling between dark
matter and dark energy.  To compare theory with
observations, we employ the Markov Chain Monte
Carlo (MCMC) methodology and use the modified
version of the program CosmoMC
\cite{Lewis:2002ah,Lewis:2013hha}, by setting the
statistical convergence for Gelman and Rubin $R -
1 = 0.03$.

The Planck data set we use is a combination of
the high-$l$ TT likelihood, which includes
measurements up to a maximum multipole number of
$l_{max} = 2500$, combined with the low-$l$ TT
likelihood which includes measurements of $l
= 2-49$
\cite{Ade:2013ktc,Ade:2013zuv,Planck:2013kta}.
Together with the Planck data, we include the
polarization measurements from the nine year
Wilkinson Microwave Anisotropy Probe (WMAP)
\cite{Bennett:2012zja}, the low-$l$ ($l<32$) TE,
EE, BB likelihood.

In addition to the CMB data sets, we also
consider Baryon Acoustic Oscillations (BAO)
measurements. We combine the results from three
data sets of BAO: the 6DF at redshift $z = 0.106$
\cite{Beutler:2011hx}, the DR7 at redshift $z = 0.35$
\cite{Padmanabhan:2012hf} and the DR9 at $z =
0.57$ \cite{Anderson:2012sa}.

Furthermore we examine the impact of the
Supernova Cosmology Project (SCP) Union 2.1
compilation \cite{Suzuki:2011hu}, which has 580
samples. Finally we also include the latest
constraint on the Hubble constant
\cite{Riess:2011yx}
\begin{equation}
H_0 = 73.8 \pm 2.4 km s^{-1}Mpc^{-1}.
\end{equation}

In a recent paper \cite{Salvatelli:2013wra}, the
authors examined the Model I of the interaction
between dark sectors listed in Table \ref{models} by
confronting to observational data including the
new measurements of the CMB anisotropies from the
Planck satellite mission. They found that the
Model I of coupled dark energy is compatible with
the Planck measurements and can relax the tension
on the Hubble constant by getting a consistent
$H_0$ as the low redshift survey such as HST and
SNIa measurements. In their analysis, they
considered ranges for the priors of different
cosmological parameters listed in Table \ref{parameters_Salvatelli} --- $\xi$
in Table \ref{parameters_Salvatelli} is the coupling constant defined in
\cite{Salvatelli:2013wra}. It relates to our
definition $\xi_2$ in Model I by dividing by $3$. At the first sight, their prior of $\Omega_c h^2$
was set unreasonably small (see note in Table \ref{parameters_Salvatelli}). It is interesting
to check, if we allow an increase of $\Omega_c
h^2$ prior, how the constraints of
cosmological parameters for Model I behave. Besides, in
\cite{Salvatelli:2013wra}, they fixed the dark
energy equation of state to be $\omega = -0.999$.
Actually there is no reason to fix the value of
$\omega$ in the global fitting. It is more
reasonable to inquire about the consequences of
setting the equation of state of dark energy to
be variable. The effect of letting $\omega$ free
to vary under the condition $\omega>-1$ was also
considered in \cite{Salvatelli:2013wra} with the priors from Table \ref{parameters_Salvatelli}. Furthermore, in
\cite{Salvatelli:2013wra}, the authors fixed the
relativistic number of degrees of freedom
parameter to $N_{eff} = 3.046$, the helium
abundance to $Y_p = 0.24$, the total neutrino
mass to $\sum m_\nu = 0.06 eV$, and the spectrum
lensing normalization to $A_L = 1$. If we change
the setting of these priors, we want to ask how
the fitting results on the Model I change. Can
Model I still be compatible with observational
data? Can the constraint on the Hubble constant
be relaxed as well? These questions are worthy of
careful study.

Besides Model I of the interaction between dark
sectors, in Table \ref{models} we have listed other three
interaction models. It would be of great interest
to carry out global fitting of these models to
the recent measurements of the CMB from the
Planck satellite mission and other complementary
observational data. In order to do so, in Table
\ref{parameters} we list the ranges for the priors of
different cosmological parameters considered in
our analysis. In our analysis we will use a big
bang nucleosynthesis (BBN) consistent scenario to
predict the primordial helium abundance $Y_p$ as
a function of the baryon density $\Omega_b h^2$
and number of extra radiation degrees of freedom
$\Delta N$. We will use interpolated results from
the PArthENoPE code \cite{Pisanti:2007hk} to set
$Y_p$, following \cite{Hamann:2011ge}.

\begin{table}[h]
\centering \caption{Initial parameters and priors
used in the analysis in \cite{Salvatelli:2013wra}
for Model I.}
\begin{tabular}{|c|c|}
\toprule
Parameters & Prior \\
\hline
$\Omega_b h^2$ & $[0.005, 0.1] $\\
\hline
$\Omega_c h^2$ & $[0.005, 0.1]$\footnote{From a
private communication, one of the authors of \cite{Salvatelli:2013wra}
told us that there was a typo in the prior of
$\Omega_c h^2$ and they used the prior for $\Omega_c
h^2$ in the range $[0.001,0.99]$.} \\
\hline
$100\theta$ & $[0.5, 10]$ \\
\hline
$\tau$ & $[0.01, 0.8]$ \\
\hline
$n_s$ & $[0.9, 1.1]$ \\
\hline
$log(10^{10} A_s)$ & $[2.7, 4]$ \\
\hline
$\xi_{2} = \xi/3$\footnote{$\xi$ defined in \cite{Salvatelli:2013wra}} & $[-0.333, 0]$ \\
\lasthline
\end{tabular}
\label{parameters_Salvatelli}
\end{table}

\begin{table}[h]
\centering \caption{The priors for cosmological
parameters considered in the analysis for
different interaction models.}
\begin{tabular}{|c|c|}
\toprule
Parameters & Prior \\
\hline
$\Omega_b h^2$ & $[0.005, 0.1] $\\
\hline
$\Omega_c h^2$ & $[0.001, 0.5]$ \\
\hline
$100\theta$ & $[0.5, 10]$ \\
\hline
$\tau$ & $[0.01, 0.8]$ \\
\hline
$n_s$ & $[0.9, 1.1]$ \\
\hline
$log(10^{10} A_s)$ & $[2.7, 4]$ \\
\hline
&
\begin{tabular}{p{1,7cm}|p{1,7cm}|p{1,7cm}|p{1,7cm}}
Model I & Model II & Model III & Model IV
\end{tabular}
\\
\hline
$\omega$ &
\begin{tabular}{p{1,7cm}|p{1,7cm}|p{1,7cm}|p{1,7cm}}
$[-1, -0.1]$ & $[-2.5, -1]$ & $[-2.5, -1]$ & $[-2.5, -1]$
\end{tabular}
\\
\hline
$\xi$ &
\begin{tabular}{p{1,7cm}|p{1,7cm}|p{1,7cm}|p{1,7cm}}
$[-0.4, 0]$ & $[0, 0.4]$ & $[0, 0.01]$ & $[0, 0.01]$
\end{tabular}
\\
\lasthline
\end{tabular}
\label{parameters}
\end{table}

\section{Fitting Results}
\label{sec:results}

We start with the Model I interacting model. We
have initially performed two runs. In the first
run we do not include the coupling, $\xi_2 = 0$,
which corresponds to the $\Lambda$CDM case, and
choose the priors of cosmological parameters
listed in Table \ref{parameters_Salvatelli}. In the second run, we follow
\cite{Salvatelli:2013wra} by setting the priors of
different cosmological parameters as in Table \ref{parameters_Salvatelli},
fixing the dark energy equation of state $\omega
= -0.999$ and setting the helium abundance $Y_p =
0.24$, the total neutrino mass $\sum m_\nu = 0.06 eV$,
and the spectrum lensing normalization $A_L= 1$.
We have let the coupling parameter $\xi_2$
to vary freely. Performing separately an analysis
with Planck data alone, we show the result in
Table \ref{bestfit_Salvatelli1}.

Our result for $\Omega_c
h^2$ obeys the prior range as indicated in Table
\ref{parameters_Salvatelli}. If we look at the Hubble constant value, in
our fitting by obeying the prior of $\Omega_c
h^2$ in Table \ref{parameters_Salvatelli}, we get higher value of $H_0$,
which shows that there is no more tension with
the Hubble Space Telescope value. But if
$\Omega_c h^2$ is above this prior range, the
$H_0$ is much smaller.
This gives us a hint that decreasing $\Omega_c
h^2$ can lead to the effect of increasing $H_0$.

The presence of a dark coupling is perfectly
compatible with the Planck data set. Our fitting
result is consistent with that shown in Table \ref{parameters_Salvatelli}
in \cite{Salvatelli:2013wra} including the value of
$H_0$ and the coupling $\xi_2$ (the relation
between our coupling and theirs is
$\xi_2=\xi/3$). While the coupled dark Model I is
compatible with most of the cosmological data, in
Table \ref{bestfit_Salvatelli1} we see that the
$\Omega_c h^2$ is unconstrained in the $1\sigma$
range although its best fitting value is still
within the set prior. This is different from the
result in Table \ref{parameters_Salvatelli} of \cite{Salvatelli:2013wra}.

We enlarge the prior to be $\Omega_c h^2 =
[0.001, 0.99]$ and perform further two runs with
Planck data alone for the $\Lambda$CDM model and
the Model I of the interacting dark sectors. We
show the results in Table
\ref{bestfit_Salvatelli2}. As expected, raising
the upper range of prior for $\Omega_c h^2$ leads
to the decrease of the values of $H_0$. This
holds for both the $\Lambda$CDM and the coupling
Model I. For the $\Lambda$CDM, our fitting result
is consistent with Table \ref{parameters_Salvatelli} in
\cite{Salvatelli:2013wra}. For coupling Model I,
we find that if we enlarge the prior of $\Omega_c
h^2$, $H_0$ is decreased, although in Table
\ref{bestfit_Salvatelli2} the fitting value of
$H_0$ is still compatible with that of HST.

In the above fittings, we followed
\cite{Salvatelli:2013wra} to fix the equation of
state of dark energy to be $\omega = -0.999$. In
the global fitting, this condition is too strong.
It is more reasonable to set the equation of
state of dark energy to be free. We choose the
prior of the equation of state of dark energy to
be in the quintessence range $\omega = [-0.999,
-0.1]$ and examine how this free parameter affects
the fitting result with Planck data alone. We
show our results in Table
\ref{bestfit_Salvatelli3}. We find that in
addition to enlarging the prior of $\Omega_c
h^2$, setting $\omega$ to be free will further
decrease the value of $H_0$ in the fitting. 
From the Planck data fitting, 
we see that the coupled dark
sectors Model I is not of much help to relax the tension of $H_0$
with the Hubble Space Telescope value.

In Tables \ref{bestfit_Salvatelli4} and
\ref{bestfit_Salvatelli5} we further show the
fitting results with Planck data alone by
fixing the helium abundance $Y_p$ to the BBN prediction and
assuming massless neutrinos, respectively. The
fitting results are basically consistent with the
result by fixing the helium abundance to $Y_p =
0.24$ and the total neutrino mass  $\sum m_\nu =
0.06 eV$, except that the constraint for the
coupling is much tighter.

We can also turn off the CMB lensing. We show the
result of fitting with Planck data alone in Table
\ref{bestfit_Salvatelli6}.  It is clear to see
that turning off the CMB lensing will further
reduce the Hubble constant at present and put
tighter constraint on the interaction.

From the above analysis, we can conclude that
although the coupled dark energy model I is fully
compatible with the Planck measurements, it is
not safe to argue that this model predicts the
Hubble constant with less tension compared with
the Hubble Space Telescope value.

\begin{table*}[tbp]
\caption{Best fit values and 68\% c.l. constraints with the parameters in Table \ref{parameters_Salvatelli}.} \centering \label{bestfit_Salvatelli1}
\begin{tabular}{ccccc}
    \toprule
    & \multicolumn{2}{c}{$\Lambda$CDM Planck} & \multicolumn{2}{c}{Interacting Planck} \\
    \cline{2-5}
    Parameter & Best fit  & 68\% limits & Best fit  & 68\% limits \\
    \hline
    $\Omega_bh^2$ & 0.02337 & $0.02330^{+0.00027}_{-0.00026}$ & 0.02197 & $0.02196^{+0.00028}_{-0.00028}$\\
    $\Omega_ch^2$ & 0.09998 & $> 0.09968$ & 0.04411 & unconstrained\\
    $H_0$ & 76.92 & $76.90^{+0.36}_{-0.37}$ & 72.93 & $72.04^{+2.26}_{-2.27}$\\
    $w$ & --- & --- & --- & ---\\
    $\xi_2$ & --- & --- & -0.1942 & $-0.1688^{+0.0732}_{-0.0713}$\\
    $\tau$ & 0.1476 & $0.1346^{+0.0170}_{-0.0189}$ & 0.09266 & $0.08751^{+0.01229}_{-0.01367}$\\
    $n_s$ & 1.013 & $1.008^{+0.005}_{-0.005}$ & 0.9607 & $0.9572^{+0.0071}_{-0.0072}$\\
    ${\rm{ln}}(10^{10}A_s)$ & 3.156 & $3.128^{+0.035}_{-0.035}$ & 3.094 & $3.083^{+0.025}_{-0.024}$\\
    \lasthline
\end{tabular}
\end{table*}

\begin{table*}[tbp]
\caption{Best fit values and 68\% c.l. constraints with the parameters in Table \ref{parameters_Salvatelli}, but with $\Omega_c h^2 = [0.001, 0.99]$} \centering \label{bestfit_Salvatelli2}
\begin{tabular}{ccccc}
    \toprule
    & \multicolumn{2}{c}{$\Lambda$CDM Planck} & \multicolumn{2}{c}{Interacting Planck} \\
    \cline{2-5}
    Parameter & Best fit  & 68\% limits & Best fit  & 68\% limits \\
    \hline
    $\Omega_bh^2$ & 0.02200 & $0.02198^{+0.000273}_{-0.000275}$ & 0.02193 & $0.02197^{+0.000278}_{-0.000277}$\\
    $\Omega_ch^2$ & 0.1195 & $0.1199^{+0.00265}_{-0.00265}$ & 0.1171 & $0.06433^{+0.0488}_{-0.0292}$\\
    $H_0$ & 67.23 & $67.1^{+1.18}_{-1.18}$ & 67.2 & $71.33^{+3.}_{-3.02}$\\
    $w$ & --- & --- & --- & ---\\
    $\xi_2$ & --- & --- & -0.009275 & $-0.1449^{+0.0837}_{-0.103}$\\
    $\tau$ & 0.08561 & $0.08868^{+0.0117}_{-0.0141}$ & 0.0923 & $0.08787^{+0.0121}_{-0.014}$\\
    $n_s$ & 0.9583 & $0.9575^{+0.00715}_{-0.00716}$ & 0.9583 & $0.9571^{+0.00701}_{-0.00722}$\\
    ${\rm{ln}}(10^{10}A_s)$ & 3.078 & $3.085^{+0.0233}_{-0.0263}$ & 3.094 & $3.084^{+0.0239}_{-0.0262}$\\
    \lasthline
\end{tabular}
\end{table*}

\begin{table*}[tbp]
\caption{Best fit values and 68\% c.l. constraints with $w = [-0.999, -0.1]$.} \centering \label{bestfit_Salvatelli3}
\begin{tabular}{ccccc}
    \toprule
    & \multicolumn{2}{c}{$\omega$CDM Planck} & \multicolumn{2}{c}{Interacting Planck} \\
    \cline{2-5}
    Parameter & Best fit  & 68\% limits & Best fit  & 68\% limits \\
    \hline
    $\Omega_bh^2$ & 0.02206 & $0.02194^{+0.00027}_{-0.00028}$ & 0.02184 & $0.02193^{+0.00027}_{-0.00027}$\\
    $\Omega_ch^2$ & 0.1180 & $0.1203^{+0.0026}_{-0.0026}$ & 0.09790 & $0.06806^{+0.04632}_{-0.02498}$\\
    $H_0$ & 65.96 & $63.09^{+4.08}_{-2.08}$ & 65.75 & $67.54^{+4.74}_{-3.22}$\\
    $w$ & -0.9348 & $<-0.8302$ & -0.9088 & $<-0.8497$\\
    $\xi_2$ & --- & --- & -0.07613 & $-0.1390^{+0.1040}_{-0.0756}$\\
    $\tau$ & 0.08683 & $0.08854^{+0.01247}_{-0.01362}$ & 0.08506 & $0.08792^{+0.01198}_{-0.01416}$\\
    $n_s$ & 0.9623 & $0.9569^{+0.0071}_{-0.0070}$ & 0.9561 & $0.9572^{+0.0073}_{-0.0072}$\\
    ${\rm{ln}}(10^{10}A_s)$ & 3.077 & $3.086^{+0.025}_{-0.024}$ & 3.084 & $3.084^{+0.023}_{-0.027}$\\
    \lasthline
\end{tabular}
\end{table*}

\begin{table*}[tbp]
\caption{Best fit values and 68\% c.l. constraints in a BBN consistency scenario.} \centering \label{bestfit_Salvatelli4}
\begin{tabular}{ccccc}
    \toprule
    & \multicolumn{2}{c}{$\omega$CDM Planck} & \multicolumn{2}{c}{Interacting Planck} \\
    \cline{2-5}
    Parameter & Best fit  & 68\% limits & Best fit  & 68\% limits \\
    \hline
    $\Omega_bh^2$ & 0.02198 & $0.02200^{+0.00028}_{-0.00029}$ & 0.02216 & $0.02202^{+0.00029}_{-0.00028}$\\
    $\Omega_ch^2$ & 0.1194 & $0.1202^{+0.0026}_{-0.0026}$ & 0.09569 & $0.06877^{+0.04806}_{-0.02449}$\\
    $H_0$ & 66.65 & $62.94^{+4.31}_{-2.28}$ & 66.75 & $67.58^{+4.98}_{-3.58}$\\
    $w$ & -0.9780 & $<-0.8176$ & -0.8946 & $<-0.8478$\\
    $\xi_2$ & --- & --- & -0.06683 & $-0.1354^{+0.1286}_{-0.0529}$\\
    $\tau$ & 0.09291 & $0.08923^{+0.01228}_{-0.01429}$ & 0.08788 & $0.08870^{+0.01220}_{-0.01410}$\\
    $n_s$ & 0.9604 & $0.9596^{+0.0071}_{-0.0072}$ & 0.9686 & $0.9600^{+0.0073}_{-0.0073}$\\
    ${\rm{ln}}(10^{10}A_s)$ & 3.096 & $3.089^{+0.024}_{-0.027}$ & 3.085 & $3.088^{+0.024}_{-0.027}$\\
    \lasthline
\end{tabular}
\end{table*}

\begin{table*}[tbp]
\caption{Best fit values and 68\% c.l. constraints with $\sum m_\nu = 0 eV$} \centering \label{bestfit_Salvatelli5}
\begin{tabular}{ccccc}
    \toprule
    & \multicolumn{2}{c}{$\omega$CDM Planck} & \multicolumn{2}{c}{Interacting Planck} \\
    \cline{2-5}
    Parameter & Best fit  & 68\% limits & Best fit  & 68\% limits \\
    \hline
    $\Omega_bh^2$ & 0.02222 & $0.02202^{+0.00028}_{-0.00028}$ & 0.02210 & $0.02203^{+0.00028}_{-0.00028}$\\
    $\Omega_ch^2$ & 0.1180 & $0.1200^{+0.0027}_{-0.0026}$ & 0.1023 & $0.07124^{+0.04748}_{-0.02382}$\\
    $H_0$ & 66.56 & $63.49^{+4.46}_{-2.26}$ & 68.10 & $67.91^{+4.88}_{-3.52}$\\
    $w$ & -0.9306 & $<-0.8177$ & -0.9480 & $<-0.8487$\\
    $\xi_2$ & --- & --- & -0.04789 & $>-0.17097$\\
    $\tau$ & 0.09347 & $0.08904^{+0.01245}_{-0.01442}$ & 0.08597 & $0.08777^{+0.01269}_{-0.01399}$\\
    $n_s$ & 0.9675 & $0.9604^{+0.0072}_{-0.0073}$ & 0.9668 & $0.9603^{+0.0073}_{-0.0073}$\\
    ${\rm{ln}}(10^{10}A_s)$ & 3.094 & $3.088^{+0.024}_{-0.027}$ & 3.082 & $3.086^{+0.025}_{-0.025}$\\
    \lasthline
\end{tabular}
\end{table*}

\begin{table*}[tbp]
\caption{Best fit values and 68\% c.l. constraints turning CMB lensing off.} \centering \label{bestfit_Salvatelli6}
\begin{tabular}{ccccc}
    \toprule
    & \multicolumn{2}{c}{$\omega$CDM Planck} & \multicolumn{2}{c}{Interacting Planck} \\
    \cline{2-5}
    Parameter & Best fit  & 68\% limits & Best fit  & 68\% limits \\
    \hline
    $\Omega_bh^2$ & 0.02021 & $0.02034^{+0.00028}_{-0.00028}$ & 0.02029 & $0.02033^{+0.00028}_{-0.00030}$\\
    $\Omega_ch^2$ & 0.1259 & $0.1254^{+0.0031}_{-0.0031}$ & 0.1072 & $0.07807^{+0.04714}_{-0.02189}$\\
    $H_0$ & 63.50 & $58.99^{+4.99}_{-2.63}$ & 63.05 & $63.73^{+5.22}_{-3.74}$\\
    $w$ & -0.9838 & $< -0.7417$ & -0.8875 & $<-0.8106$\\
    $\xi_2$ & --- & --- & -0.06078 & $>-0.19240$\\
    $\tau$ & 0.07279 & $0.07643^{+0.01115}_{-0.01260}$ & 0.06450 & $0.07622^{+0.01121}_{-0.01275}$\\
    $n_s$ & 0.9358 & $0.9339^{+0.0076}_{-0.0083}$ & 0.9324 & $0.9337^{+0.0077}_{-0.0078}$\\
    ${\rm{ln}}(10^{10}A_s)$ & 3.059 & $3.065^{+0.022}_{-0.025}$ & 3.036 & $3.065^{+0.023}_{-0.026}$\\
    \lasthline
\end{tabular}
\end{table*}

Besides the interacting dark sector Model I, we
would like to put constraints on other coupled
dark energy models listed in Table \ref{models} from the
recent measurements of the Cosmic Microwave
Background Anisotropies from the Planck satellite
mission. We will also consider the combined
constraints for the general phenomenological
interacting models between dark sectors from the
Planck data plus the BAO measurements, SNIa and
HST observational data. In our analysis, we will
choose our priors of different cosmological
parameters as listed in Table \ref{parameters}. We will allow
the equation of state of dark energy to vary and
choose the helium abundance $Y_p$ from a BBN
consistency scenario. We will take the relativistic number of degrees of freedom $N_{eff} = 3.046$, the total neutrino mass to $\sum m_\nu = 0.06 eV$ and the spectrum
lensing normalization to $A_L = 1$. After
running the MCMC, we list our fitting results in
Tables \ref{bestfit1}-\ref{bestfit4}.

\begin{figure}[H]
\centering
\includegraphics[scale=0.45]{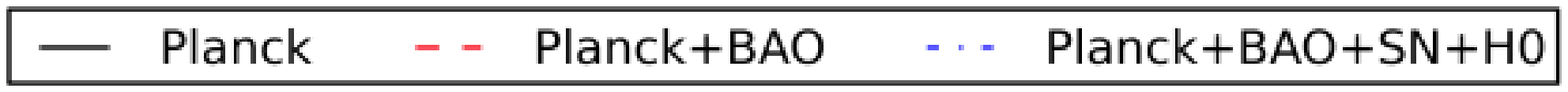}
\includegraphics[scale=0.45]{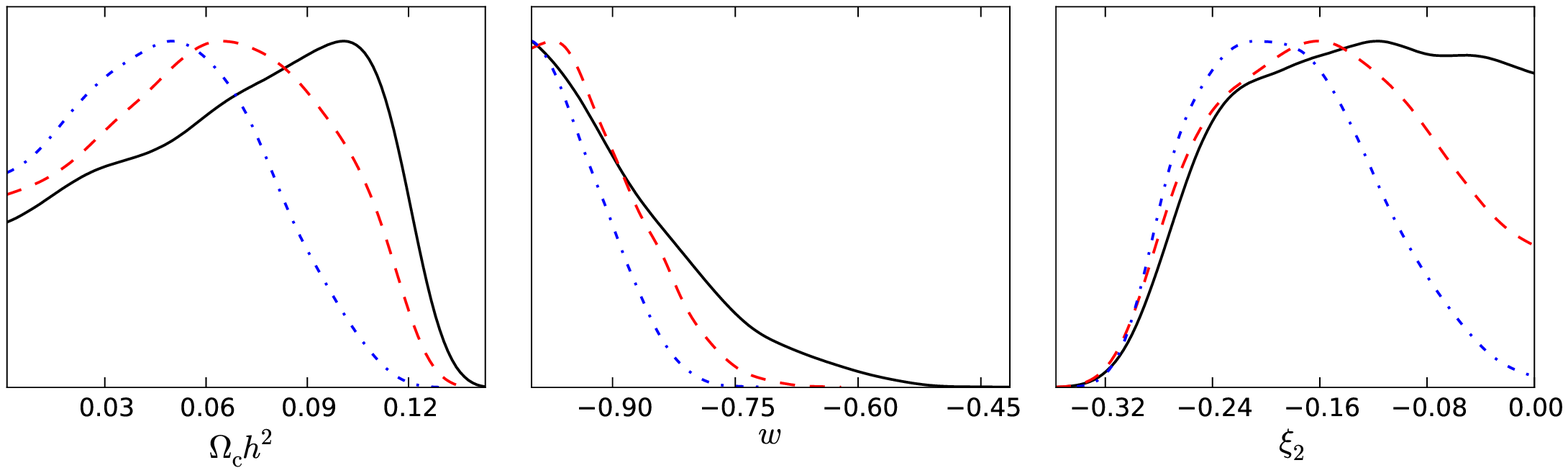}
\includegraphics[scale=0.45]{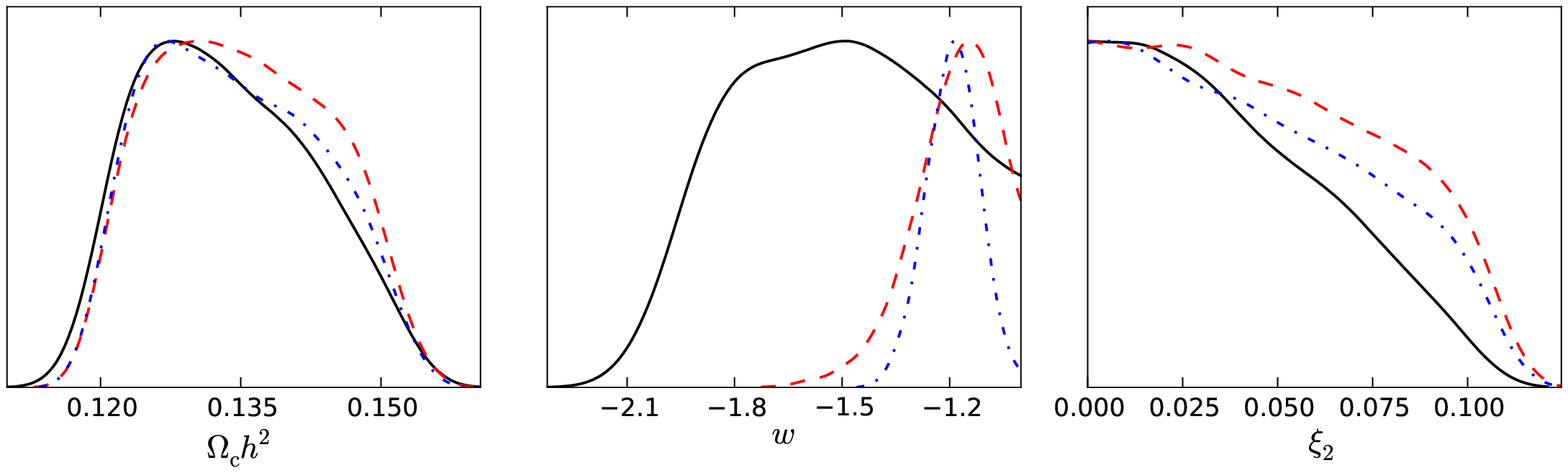}
\includegraphics[scale=0.45]{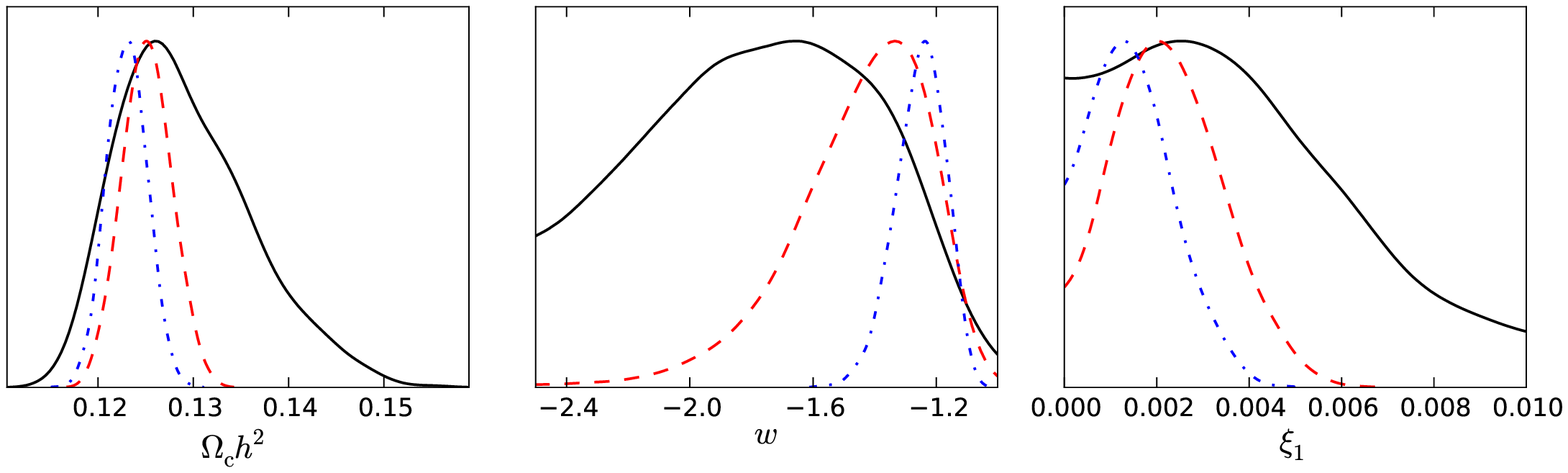}
\includegraphics[scale=0.45]{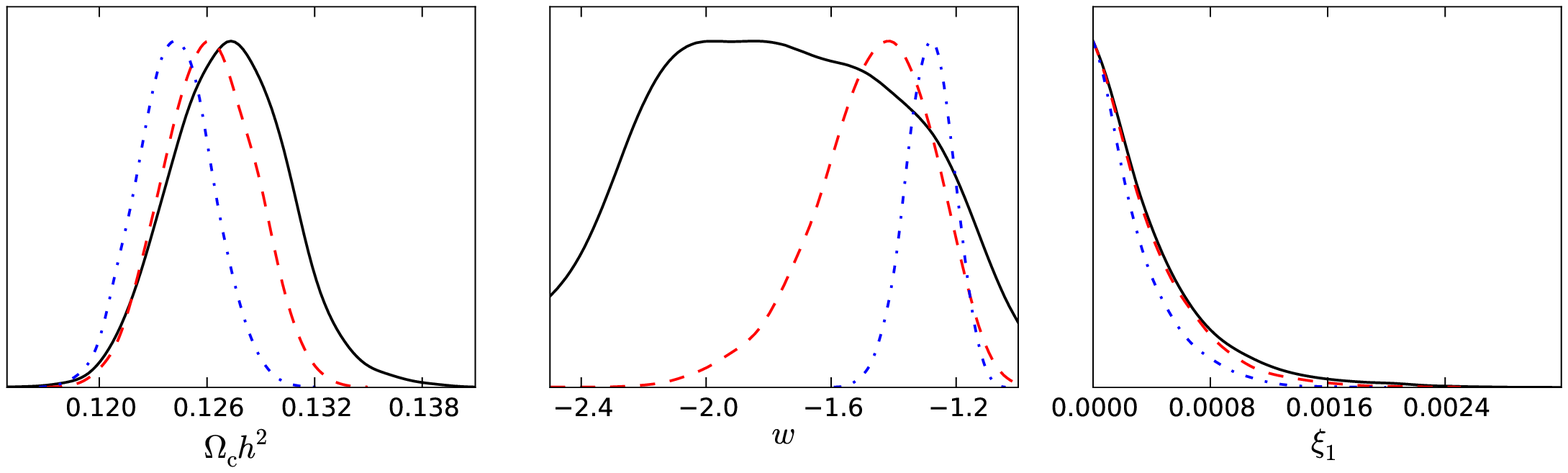}
\caption{(Color online). The likelihood of cold dark matter abundance $\Omega_c h^2$, dark energy EoS $\omega$ and couplings $\xi$ for the four models.}
\label{1d_distribution}
\end{figure}

\begin{figure}[H]
    \includegraphics[width=\textwidth]{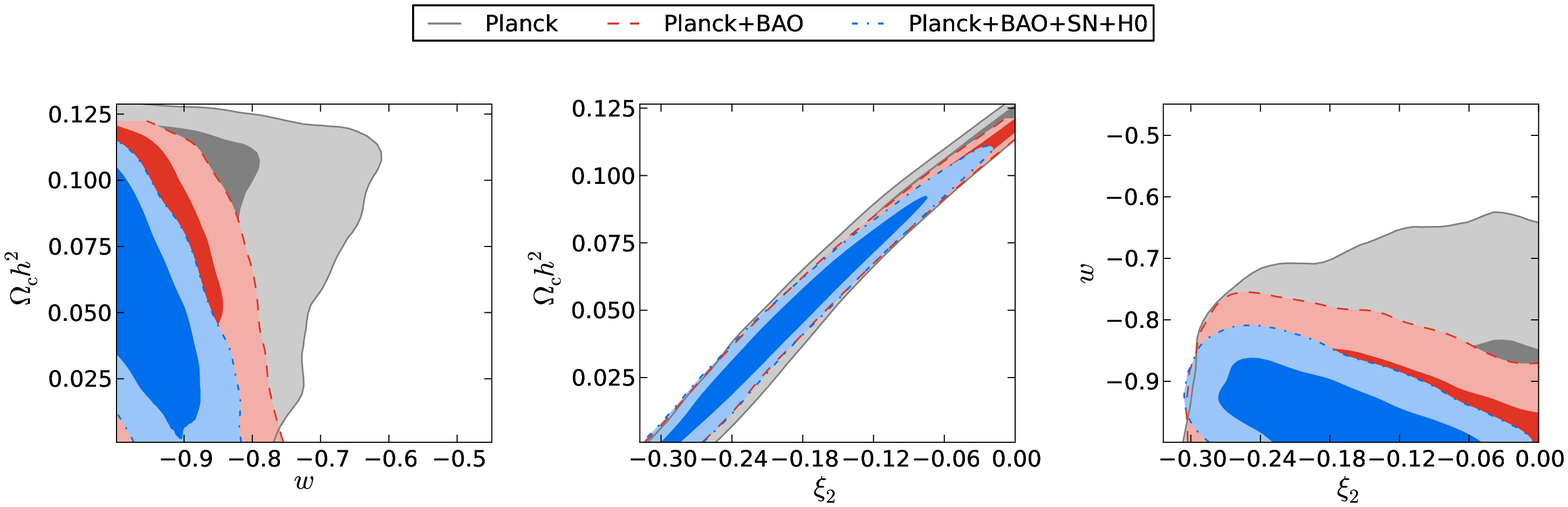}
\caption{(Color online). 2-D distribution for selected parameters - Model I.}\label{2d_dist1}
\end{figure}

The constraints on the parameters and the best
fit values for Model I are reported in Table
\ref{bestfit1}. The 1-D posteriors for the
parameters $\Omega_c h^2$, $\omega$ and $\xi_2$
are shown at the top row of Fig.\ref{1d_distribution} and the main
parameter degeneracies are shown in Fig.\ref{2d_dist1}. The
presence of a dark coupling is perfectly
compatible with the Planck data set. The
marginalized value tells us $\xi_2<0$. With the
combined constraint by including other
observational data, the negative value of the
coupling keeps, which shows that in this coupling
model, there is a lower value of the cold dark
matter density today, since there is energy flow
from dark matter to dark energy. This direction
of energy flow cannot alleviate the coincidence.
As shown in Fig.\ref{ratio_rho}, there is even shorter period
for the energy densities of dark matter and dark
energy to be comparable. For the Hubble constant
value, from the Planck data alone, $H_0$ is small
in this interacting model, which is similar to
that obtained in the $\Lambda$CDM case. This
interaction model between dark sectors cannot be
of much help to relax the tension on the Hubble
parameter between Planck measurement and HST
observation. After including other observational
data at low redshift, we find that the tension
between the Hubble constant measurements is
alleviated.

\begin{figure}[H]
    \includegraphics[width=\textwidth]{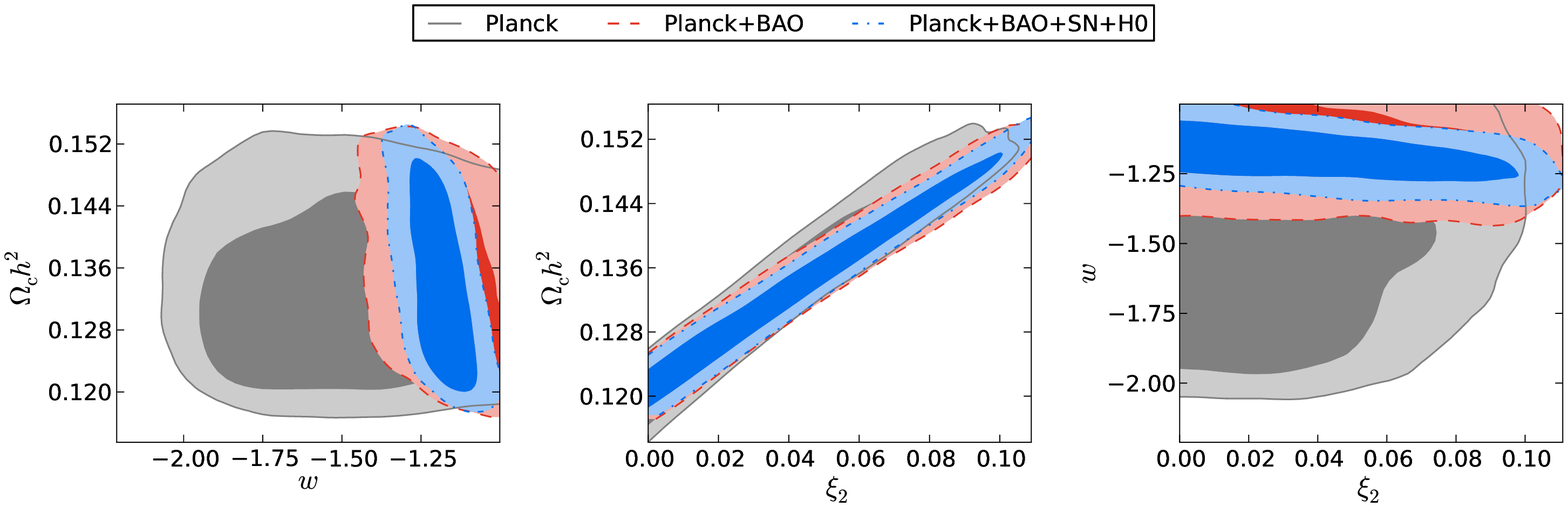}
\caption{(Color online). 2-D distribution for selected parameters - Model II.}\label{2d_dist2}
\end{figure}

Now we present the fitting result for the
coupling Model II in Table \ref{bestfit2}, where
the interaction between dark sectors is still
proportional to the energy density of dark energy
but with equation of state of dark energy smaller
than $-1$. From the Planck data analysis alone,
for this coupled dark energy model, using our
cosmological parameters prior listed in Table
\ref{parameters}, we obtain the Hubble constant value
significantly larger than that in the standard
$\Lambda$CDM case, $H_0 =
82.69^{+9.78}_{-11.9}~\mathrm{km \cdot s^{-1}
\cdot Mpc^{-1}}$. This is different from what we
observed in the fitting result of Model I, where
the $H_0$ is much smaller and consistent with the
$\Lambda$CDM case. The lower fitting range of the
$H_0$ in Model II is consistent with the
observations in the low redshift. We have
explored the degeneracy between the Hubble value
and the equation of state of dark energy and
found that smaller equation of state of dark
energy leads to higher value of the Hubble
parameter. The coupling constant $\xi_2$ is found
to be positive, which shows that there is an
energy flow from dark energy to dark matter. This
is required to alleviate the coincidence problem,
because with this interaction there is longer
period for the energy densities of dark matter
and dark energy to be comparable, which was
illustrated in the Fig.\ref{ratio_rho}. Combined with other
observational data, we show that a combined
analysis provides significant evidence for this
coupled dark energy with positive non-zero value
of the coupling parameter, consistent Hubble
constant and equation of state of dark energy.
The 1-D posteriors for the parameters $\Omega_c
h^2$, $\omega$ and $\xi_2$ are shown in the
second row of Fig.\ref{1d_distribution} and the main parameter
degeneracies are shown in Fig.\ref{2d_dist2}.

\begin{figure}[H]
    \includegraphics[width=\textwidth]{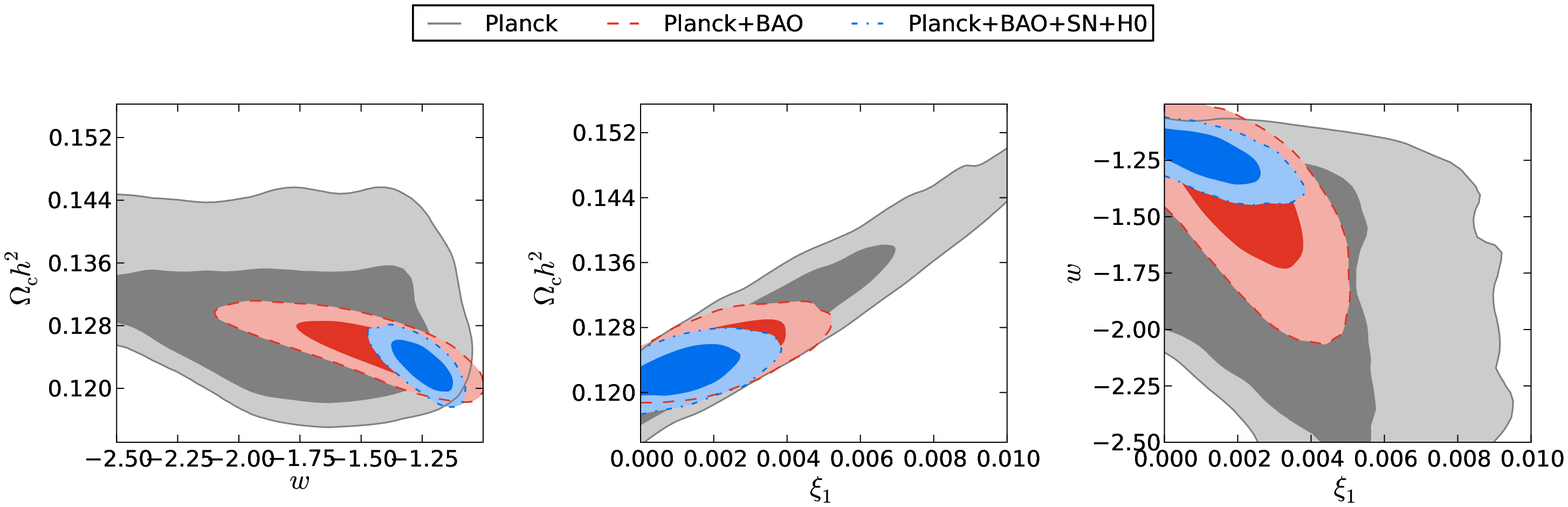}
\caption{(Color online). 2-D distribution for selected parameters - Model III.}\label{2d_dist3}
\end{figure}

Now we turn our discussion to the coupled dark
energy Model III, where the interaction is
proportional to the energy density of dark
matter. To ensure stability of the curvature
perturbation, in this model if the equation of
state of dark energy is constant, it has to be
smaller than $-1$ \cite{He:2008si}. Looking at
the new constraints on this coupled dark energy
model from the recent measurements of CMB from
the Planck satellite mission alone in Table
\ref{bestfit3}, we find that the Hubble constant
value is consistent with low redshift
observations, but it is much higher than that of
the $\Lambda$CDM result. The coupling constant is
more tightly constrained in this coupled dark
energy model than those in Models I and II, which
is in agreement with the findings in the WMAP
constraints \cite{He:2010im,He:2009pd}. The value
of the coupling parameter $\xi_1$ is small
positive, which meets the requirement to
alleviate the coincidence problem. The evolution
of the ratio between energy densities of dark
matter and dark energy with this small positive
coupling was shown in the Fig.\ref{ratio_rho}, which has a
longer period for the dark matter and dark energy
energy densities to be comparable when $\xi$ is
positive and has the attractor solution with the
ratio between dark energy and dark matter energy
densities $r\sim constant$ in the past. We also
consider the combined constraints from the Planck
data plus other measurements. The results are
listed in Table \ref{bestfit3}, which shows
stronger evidence for this coupled dark energy
model with small positive coupling. We plot the
1-D posteriors for the parameters $\Omega_c h^2$,
$\omega$ and $\xi$ in the third row of Fig.\ref{1d_distribution} and
show the main parameter degeneracies in Fig.\ref{2d_dist3}.

\begin{figure}[H]
    \includegraphics[width=\textwidth]{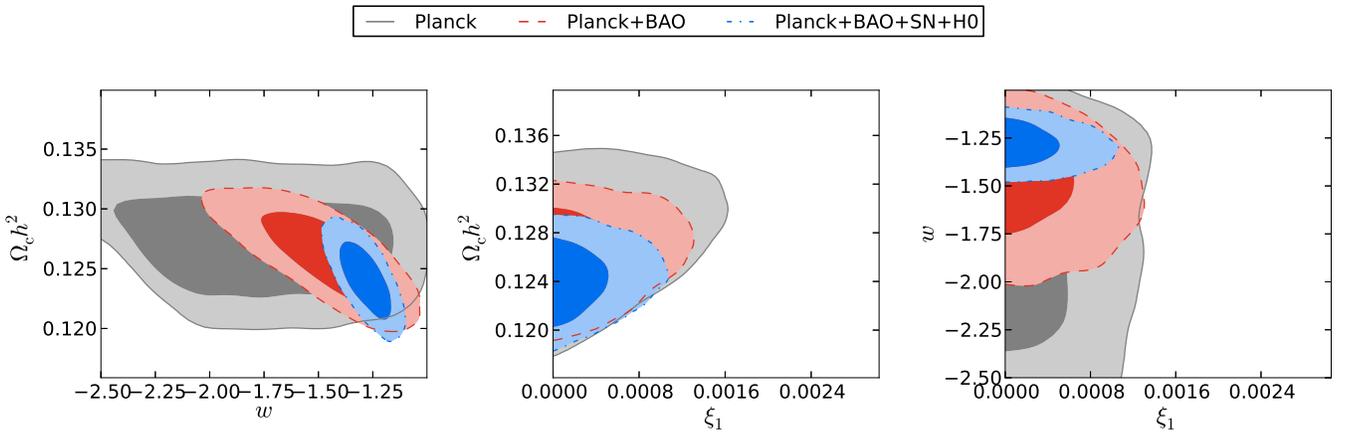}
\caption{(Color online). 2-D distribution for selected parameters - Model IV.}\label{2d_dist4}
\end{figure}

Finally we present the fitting results for the
coupled dark energy Model IV, where we consider
the interaction between dark energy and dark
matter is proportional to the energy density of
the total dark sectors. In order to ensure the
stability of the curvature perturbation, for the
constant equation of state of dark energy, it has
to be in the phantom range. This was disclosed in
\cite{He:2008si}. As observed in the WMAP fitting
results, this type of interaction has very
similar constraints to the Model III
\cite{He:2010im,He:2009pd}. Confronting the model
to the Planck data alone and the combined
observational data, we list the constraints in
Table \ref{bestfit4}. We show the 1-D posteriors
for the parameters $\Omega_c h^2$, $\omega$ and
$\xi$ in the fourth row of Fig.\ref{1d_distribution} and plot the
main parameter degeneracies in Fig.\ref{2d_dist4}. From the
Planck data alone, we again see that for this
interacting dark energy model, the Hubble
constant is much higher than that of the
$\Lambda$CDM model. This is consistent with the
observations from Model II and Model III. The
coupling constant is more tightly constrained in
Model IV to be very small but positive, what is
needed to alleviate the coincidence problem with
longer period for the dark energy and dark matter
energy densities to be comparable in the expansion
of the universe as shown in Fig.\ref{ratio_rho}. The Model IV
has an attractor solution with $r\sim constant$
in the future. In the joint constraints, by
including other observational data, we find that
the coupled dark energy model IV is fully
compatible with astronomical observations. It is
a viable model.

\begin{table*}[tbp]
\caption{Cosmological parameters - Model I.} \centering \label{bestfit1}
\begin{tabular}{ccccccc}
    \toprule
    & \multicolumn{2}{c}{Planck} & \multicolumn{2}{c}{Planck+BAO} & \multicolumn{2}{c}{Planck+BAO+SNIa+H0} \\
    \cline{2-7}
    Parameter & Best fit  & 68\% limits & Best fit  & 68\% limits & Best fit  & 68\% limits \\
    \hline
$\Omega_b h^2$ & 0.02213 & $0.02202^{+0.000272}_{-0.000273}$ & 0.02225 & $0.02203^{+0.000261}_{-0.000261}$ & 0.0221 & $0.02202^{+0.000251}_{-0.000251}$\\
$\Omega_c h^2$ & 0.1188 & $0.06889^{+0.0483}_{-0.0252}$ & 0.1121 & $0.0608^{+0.038}_{-0.0311}$ & 0.07199 & $0.04824^{+0.0256}_{-0.0319}$\\
$H_0$ & 66.81 & $67.66^{+4.7}_{-3.55}$ & 68.26 & $69.26^{+2.04}_{-1.99}$ & 70.72 & $70.71^{+1.36}_{-1.37}$\\
$w$ & -0.9747 & $-0.8797^{+0.0287}_{-0.119}$ & -0.9934 & $-0.9141^{+0.0221}_{-0.0849}$ & -0.9935 & $-0.9362^{+0.0171}_{-0.0628}$\\
$\xi_2$ & -0.0006633 & $-0.1353^{+0.128}_{-0.0528}$ & -0.02123 & $-0.1546^{+0.0743}_{-0.0947}$ & -0.1359 & $-0.1854^{+0.0524}_{-0.0793}$\\
$\tau$ & 0.08951 & $0.08843^{+0.0123}_{-0.0136}$ & 0.09803 & $0.08835^{+0.0121}_{-0.0139}$ & 0.09492 & $0.08866^{+0.012}_{-0.0136}$\\
$n_s$ & 0.9596 & $0.9601^{+0.00747}_{-0.00739}$ & 0.9643 & $0.9606^{+0.00639}_{-0.00642}$ & 0.964 & $0.9598^{+0.00616}_{-0.00624}$\\
${\rm{ln}}(10^{10} A_s)$ & 3.088 & $3.087^{+0.0237}_{-0.0256}$ & 3.106 & $3.086^{+0.0238}_{-0.0265}$ & 3.102 & $3.088^{+0.0236}_{-0.0261}$\\
    \lasthline
\end{tabular}
\end{table*}

\begin{table*}[tbp]
\caption{Cosmological parameters - Model II.} \centering \label{bestfit2}
\begin{tabular}{ccccccc}
    \toprule
    & \multicolumn{2}{c}{Planck} & \multicolumn{2}{c}{Planck+BAO} & \multicolumn{2}{c}{Planck+BAO+SNIa+H0} \\
    \cline{2-7}
    Parameter & Best fit  & 68\% limits & Best fit  & 68\% limits & Best fit  & 68\% limits \\
    \hline
$\Omega_b h^2$ & 0.02201 & $0.02208^{+0.000283}_{-0.000277}$ & 0.02219 & $0.02199^{+0.000264}_{-0.00026}$ & 0.02208 & $0.02203^{+0.000255}_{-0.000255}$\\
$\Omega_c h^2$ & 0.1308 & $0.1335^{+0.0076}_{-0.0118}$ & 0.132 & $0.1352^{+0.00844}_{-0.0115}$ & 0.1432 & $0.1344^{+0.00751}_{-0.0118}$\\
$H_0$ & 88.93 & $82.69^{+9.78}_{-11.9}$ & 70.68 & $70.92^{+2.08}_{-3.19}$ & 70.42 & $71.25^{+1.48}_{-1.48}$\\
$w$ & -1.696 & $-1.516^{+0.312}_{-0.305}$ & -1.166 & $-1.189^{+0.152}_{-0.0721}$ & -1.181 & $-1.192^{+0.0771}_{-0.0715}$\\
$\xi_2$ & 0.02837 & $0.03923^{+0.0121}_{-0.0392}$ & 0.03522 & $0.04818^{+0.0164}_{-0.0482}$ & 0.0784 & $0.04562^{+0.0155}_{-0.0456}$\\
$\tau$ & 0.08672 & $0.08934^{+0.0128}_{-0.0138}$ & 0.08154 & $0.08761^{+0.0121}_{-0.0137}$ & 0.08312 & $0.08844^{+0.012}_{-0.0135}$\\
$n_s$ & 0.9615 & $0.9599^{+0.00715}_{-0.00703}$ & 0.9598 & $0.9581^{+0.00654}_{-0.00658}$ & 0.962 & $0.9586^{+0.00632}_{-0.00637}$\\
${\rm{ln}}(10^{10} A_s)$ & 3.085 & $3.089^{+0.0245}_{-0.0267}$ & 3.078 & $3.088^{+0.0234}_{-0.0261}$ & 3.079 & $3.089^{+0.0232}_{-0.0263}$\\
    \lasthline
\end{tabular}
\end{table*}

\begin{table*}[tbp]
\caption{Cosmological parameters - Model III.} \centering \label{bestfit3}
\begin{tabular}{ccccccc}
    \toprule
    & \multicolumn{2}{c}{Planck} & \multicolumn{2}{c}{Planck+BAO} & \multicolumn{2}{c}{Planck+BAO+SNIa+H0} \\
    \cline{2-7}
    Parameter & Best fit  & 68\% limits & Best fit  & 68\% limits & Best fit  & 68\% limits \\
    \hline
    $\Omega_b h^2$ & 0.02225 & $0.02265^{+0.000412}_{-0.000506}$ & 0.02248 & $0.02244^{+0.000347}_{-0.000399}$ & 0.02227 & $0.02235^{+0.000314}_{-0.000372}$\\
    $\Omega_c h^2$ & 0.1258 & $0.1292^{+0.00516}_{-0.00857}$ & 0.1254 & $0.1251^{+0.00256}_{-0.00257}$ & 0.1237 & $0.123^{+0.00212}_{-0.00212}$\\
    $H_0$ & 79.85 & $79.35^{+12.4}_{-12.1}$ & 76.02 & $75.23^{+2.73}_{-4.91}$ & 72.24 & $71.88^{+1.44}_{-1.43}$\\
    $w$ & -1.638 & $-1.779^{+0.457}_{-0.341}$ & -1.48 & $-1.455^{+0.275}_{-0.139}$ & -1.296 & $-1.254^{+0.0944}_{-0.0695}$\\
    $\xi_1$ & 0.002118 & $<0.004702$ & 0.002266 & $0.002272^{+0.00103}_{-0.00137}$ & 0.001781 & $0.001494^{+0.00065}_{-0.00116}$\\
    $\tau$ & 0.08378 & $0.08887^{+0.013}_{-0.0131}$ & 0.09507 & $0.08956^{+0.0126}_{-0.0142}$ & 0.08342 & $0.09011^{+0.0124}_{-0.0141}$\\
    $n_s$ & 0.9584 & $0.9563^{+0.00756}_{-0.00758}$ & 0.9603 & $0.9587^{+0.00651}_{-0.00667}$ & 0.9631 & $0.9599^{+0.00614}_{-0.0062}$\\
    ${\rm{ln}}(10^{10}A_s)$ & 3.075 & $3.081^{+0.0252}_{-0.0269}$ & 3.095 & $3.084^{+0.0246}_{-0.0269}$ & 3.071 & $3.086^{+0.0239}_{-0.0273}$\\
    \lasthline
\end{tabular}
\end{table*}

\begin{table*}[tbp]
\caption{Cosmological parameters - Model IV.} \centering \label{bestfit4}
\begin{tabular}{ccccccc}
    \toprule
    & \multicolumn{2}{c}{Planck} & \multicolumn{2}{c}{Planck+BAO} & \multicolumn{2}{c}{Planck+BAO+SNIa+H0} \\
    \cline{2-7}
    Parameter & Best fit  & 68\% limits & Best fit  & 68\% limits & Best fit  & 68\% limits \\
    \hline
    $\Omega_b h^2$ & 0.02047 & $0.02037^{+0.000275}_{-0.00027}$ & 0.02041 & $0.02042^{+0.000257}_{-0.000263}$ & 0.02053 & $0.02056^{+0.000253}_{-0.000265}$\\
    $\Omega_c h^2$ & 0.1251 & $0.1273^{+0.00309}_{-0.00321}$ & 0.125 & $0.1261^{+0.00254}_{-0.0025}$ & 0.1245 & $0.1242^{+0.00204}_{-0.00208}$\\
    $H_0$ & 80.35 & $82.5^{+12.4}_{-9.95}$ & 70.71 & $75.^{+3.07}_{-4.59}$ & 72.11 & $71.45^{+1.48}_{-1.46}$\\
    $w$ & -1.613 & $-1.763^{+0.385}_{-0.432}$ & -1.267 & $-1.472^{+0.229}_{-0.147}$ & -1.305 & $-1.286^{+0.082}_{-0.074}$\\
    $\xi_1$ & 0.00009881 & $<0.0004618$ & 0.00001943 & $<0.0004260$ & 0.0000671 & $<0.0003314$\\
    $\tau$ & 0.0883 & $0.07771^{+0.011}_{-0.0129}$ & 0.06756 & $0.07785^{+0.0112}_{-0.0124}$ & 0.07537 & $0.07899^{+0.0112}_{-0.0127}$\\
    $n_s$ & 0.9305 & $0.9309^{+0.00746}_{-0.00743}$ & 0.9295 & $0.9332^{+0.00643}_{-0.00655}$ & 0.9338 & $0.9368^{+0.00592}_{-0.00594}$\\
    ${\rm{ln}}(10^{10}A_s)$ & 3.086 & $3.068^{+0.0221}_{-0.0253}$ & 3.045 & $3.066^{+0.0228}_{-0.0248}$ & 3.06 & $3.064^{+0.0227}_{-0.0233}$\\
    \lasthline
\end{tabular}
\end{table*}

\section{Conclusions}

In this paper we have presented cosmological
constraints on general phenomenological dark
matter-dark energy interaction models from the
new CMB measurements provided by the Planck
experiment. We have found that a dark coupling
interaction is compatible with Planck data. For
Model I, the coupling parameter is weakly
constrained  to  negative  values by Planck
measurements, while for the other three models
the coupling constants are all positive from
Planck data constraints. The positive coupling
indicating that there is energy flow from dark
energy to dark matter, as required to alleviate
the coincidence problem and to satisfy the second
law of thermodynamics \cite{Pavon:2007gt}. Thus
Model II, III and IV are very reassuring in the
light of the coincidence problem.

It was claimed that Model I gives a larger Hubble
parameter compatible with the HST value
\cite{Salvatelli:2013wra}. However, this heavily
depends on the prior of $\Omega_c h^2$,  the
fixed value of $\omega$ they chose and other
factors. If we enlarge the prior of $\Omega_c
h^2$ and allow $\omega$ to vary in the
quintessence range, the $H_0$ constrained in
Model I can be lower than the HST value and is
consistent with the value in the $\Lambda$CDM
case.  Thus, the coupled dark energy Model I
cannot be counted to resolve the tension between
the Planck and the HST measurements of the Hubble
parameter.

After examining the fitting results for the other
phenomenological coupled dark energy models, we
find that the dark interaction in Models II, III
and IV can give a larger Hubble parameter. There
is degeneracy between the Hubble parameter and
the equation of state of dark energy. If future
data can constrain $\omega$ closer to $-1$ from
below, the fitting result of the Hubble parameter
can be more consistent with the HST value. Thus
Models II, III and IV have the possibility to
relax the tension of the Hubble parameter between
the Planck and the HST measurements.

We have also considered the combined constraints
from the Planck data plus other observations.
These analyzes have provided significant evidence
that the phenomenological coupled dark energy
models are viable. Taking into account all data
sets, it appears in the data fittings that Model
I shows the most significant departure from zero
coupling, although it does not help to alleviate
the coincidence problem.

The weak point of these models is the fact that
the equation of state is fixed, not depending on
time. In a more realistic model, we expect it to
be time dependent (or else, redshift dependent).
In order to probe such a statement we need a
model grounded on cosmological fields rather than
on simple phenomenology, e.g. coupled
quintessence models \cite{Pettorino:2013oxa}.
This is currently under investigation.

\begin{acknowledgments}

This work was partially supported by the National
basic research program of China (2013CB834900)
and National Science Foundation of China. A.C.,
E.A. and E.F. acknowledges financial support from
CNPq (Conselho Nacional de Desenvolvimento
Cient\'\i fico e Tecnol\' ogico), and E.A. and
E.F. also from FAPESP (Funda\c c\~ao de Amparo
\`a Pesquisa do Estado de S\~ao Paulo).

This work has made use of the computing facilities of the Laboratory of Astroinformatics (IAG/USP, NAT/Unicsul), whose purchase was made possible by the Brazilian agency FAPESP (grant 2009/54006-4) and the INCT-A.

\end{acknowledgments}

\bibliographystyle{apsrev4-1}
\bibliography{references}

\end{document}